\documentclass[pra,floatfix,preprint,nofootinbib,12pt]{revtex4}
 \usepackage{amsmath}
\def\cN{\cal N} 

\begin{document}
 
 %\centerline{\Huge  DRAFT 3 \today}
 
\title{The Impact of Cosmology on Quantum Mechanics\footnote{A pedagogical essay. Based in part on a talk given at the conference {\it 90 Years of Quantum Mechanics}. Singapore, January, 2017}}
\author{James B.~Hartle}

\email{hartle@physics.ucsb.edu}

\affiliation{Department of Physics, University of California,Santa Barbara, CA 93106-9530}
\affiliation{Santa Fe Institute, Santa Fe, NM 87501}

 \title{Quantum Buzzwords\footnote{A pedagogical essay that is based on a modestly reworked and expanded  appendix from the author's  {\it The Quantum Mechanics of Cosmology }in  {\it Quantum Cosmology and Baby Universes: Proceedings of the 1989
Jerusalem Winter School for Theoretical Physics}  (edited by  S. ~Coleman, J.~Hartle, 
T. Piran, and S. Weinberg), World Scientific, Singapore (1991).  } }
 
 \author{James B.~Hartle}

\email{hartle@physics.ucsb.edu}

\affiliation{Santa Fe Institute, Santa Fe, NM 87501}
\affiliation{Department of Physics, University of California,
 Santa Barbara, CA 93106-9530}
\begin{abstract} 
%\singlespacing
Many scientists seeking  to understand  the quantum mechanics of measurement situations  (Copenhagen quantum theory) agree on its overwhelmingly successful 
algorithms to predict the outcomes of laboratory measurements but disagree on what these algorithms mean and how they are to be interpreted.  Some of these problems are briefly described and resolutions suggested from the  decoherent (or consistent) histories   quantum mechanics of closed systems like the Universe.
\end{abstract}

\maketitle

\tableofcontents

{\vskip -1in}

%\tableofcontents

%documentclass[12pt,nofootinbib,tightenlines]{revtex4} 

%\def\vs{\vskip .2in} 

% \vskip .1in

%\centerline{James. B. Hartle}
%\vskip .1in 

%\vskip .2in

\eject

\section{Introduction}
\label{intro}
The Copenhagen quantum mechanics   of measurement situations  (CQM) found in standard textbooks\footnote{By `Copenhagen quantum mechanics', or `textbook quantum mechanics'  we mean the standard formulation that is found in many textbooks. We do not necessarily mean that it conforms exactly to what the founders of quantum mechanics meant by the `Copenhagen interpretation of quantum mechanics' \cite{Jam66}.} is arguably the most successful theoretical framework in the history of physics. It is central to our understanding of a vast range of physical phenomena, including atoms, molecules, chemistry, the solid state, how stars are formed, shine, evolve, and die, nuclear energy, thermonuclear explosions, how transistors work, and many, many, many, more phenomena. It has  been claimed that quantum mechanics is responsible for a significant fraction of the US GDP. 

Work continues today to better understand Copenhagen quantum mechanics, to make its central notions of measurement and state vector reduction more precise, and to resolve `problems' that it is alleged to have.  Despite the lack of such precision  early in its history, the author knows of no mistakes that were made as a consequence  in correctly applying  CQM in the century since it was first formulated. 

A few `quantum buzzwords' (or phrases) characterize some of the issues with CQM that challenge understanding.  A short list would include  `the definition of measurement', `state vector collapse', `many worlds' , `the locality of quantum theory', `quantum states of subsystems', `Schr\"odinger's cat', `living in a superposition' , `reality,' the `quantum arrow of time',' consciousness',  `the Heisenberg cut' , `observers' `a role for consciousness',  `states for subsystems', the principle of superposition, ... There are undoubtably others.\footnote{Some topics that require a discussion that would be too extensive for this paper have been left out. The `quantum measurement problem' is one, but see  e.g. \cite{Wal12}. Another is the
 many ways of deriving Born's rule from the other postulates of quantum mechanics.  See \cite{Har68} for the author's effort on this. There are a great many others .}

The Copenhagen quantum mechanics of measurement situations  has to generalized for several reasons. The simplest reason is that quantum phenomena are not restricted just to laboratory measurements. Much of what we observe  of our large scale universe is a consequence of quantum processes that started in the very early universe.\footnote{A well known example is the growth of quantum density fluctuations in the early universe and their subsequent collapse due to gravitational attraction. We see the results in the fluctuations in the cosmic background radiation (CMB) a few hundred thousand years after the big bang and in the distribution of the galaxies in the present.}  A generalization of CQM is thus needed for cosmology \cite{Har19}.

CQM must also be generalized is to contribute to  the current  quest for  `final theories' that unifiy all interactions including gravity. CQM assumes a fixed, flat, classical spacetime. But in the very early universe, near the big-bang, spacetime will neither be flat nor fixed,. Rather it will be fluctuating quantum mechanically. A generalization of CQM is needed just to retrodict the probabilities of what went on then.

 Generalizations of CQM can shed light on issues labeled by the quantum buzzwords.
This paper is devoted to  suggestions of how the consistent, or decoherent histories formulation of quantum theory (DHQM)  \cite{Gri84,classicDH} could help understand and resolve some of these concerns\footnote{Robert Griffiths introduced consistent histories quantum mechanics in \cite{Gri84}.  Independently, but later,  the author  and Murray Gell-Mann introduced a closely related framework which they called decoherent histories quantum theory \cite{GH90}.  Consistent histories quantum mechanics  and decoherent histories quantum mechanics may be distinguished by their historical development, by  different emphases, by terminology, and by application.  But, in the author's opinion, they both denote essentially the same framework for quantum prediction.  In this paper we use the term `decoherent histories' to be compatible with the author's other writings.} 

DHQM  is a generalization of CQM that applies to closed systems like a spatially closed universe. It is an extension,  and to some extent a completion, of the ideas pioneered by Everertt \cite{Eve57}.  Given theories of the closed system's quantum state and dynamics, DHQM  predicts probabilities for the individual members of sets of alternative coarse-grained  time histories of what goes on in the closed system.  These could be alternative histories of the expansion of the universe in cosmology for example. They could also be histories of how the universe began in the big bang.  But they can also be alternative histories of the preparation and outcomes of an experiment in a laboratory along with the behavior of observers who are carrying out the experiment.  CQM can  thus be  seen as an emergent approximation to the DHQM  quantum mechanics of a closed system like the universe.. That  viewpoint  gives a different and  fruitful perspective on the quantum buzzwords. 

Section \ref{back}  contains a bare bones exposition of  the parts of DHQM that we will need for our discussion. Section III contains a discussion of these buzzwords. Section \ref{conc} offers some brief conclusions and  suggests directions for further research.  An appendix 
contains an unpublished manuscript by Murray Gell-Mann and the author entitled {\it `Copenhagen Quantum Mechanics and Decoherent Histories Quantum Mechanics} \cite{GH06}. It covers much the same ground as this paper but more briefly,  with a different voice, and different emphases that may be useful.

 This paper is  not an in depth review or an exposition of either CQM or DHQM and their applications.  And it is certainly not of a history of the now century old discussion  of these topics.  The author aims only to provide a few brief suggestions for understanding these issues presented by the buzzwords and some references to follow these suggestions up. In one way or another most of the issues are discussed in the author's papers and essays and therefore  many of the references are to these. No models are developed in this paper and no calculations are presented.  We aim rather at brief discussions of that might motivate models and calculation.  Discussions at greater depth (and greater length) can be found in the author's collection of essays\footnote{ ``The Quantum Universe'', World Scientific, Singapore or arXiv1807.04126}.   For the most part the discussion is about  the facts of these formulations and their connection. But occasionally author's opinions on how to proceed are offered.

\section{ Decoherent Histories Quantum Mechanics. (DHQM) \\
\indent A Brief Introduction} 
\label{back}
\begin{subsection}{Setup}
\label{setup}
To simplify the discussion  we assume throughout a flat classical background spacetime with which to give time histories meaning.  We are thus neglecting quantum gravity for the most part\footnote{For more on DHQM and quantum spacetime  see e.g. \cite{Jer}.}.
 We consider a closed quantum system with a Hilbert space $\cal H$. A spatially closed universe is a relevant  example. The basic theoretical inputs to prediction are assumed to be  a pure quantum state of the universe 
$|\Psi\rangle$    and a Hamiltonian $H$ specifying the system's quantum dynamics.  In the Heisenberg picture in which we work  neither varies in time.  We aim at predicting the probabilities of  of the individual members of sets of alternative, coarse-grained   time histories of  this system.  Theories of  $|\Psi\rangle$ and  $H$, together with  the assumed  classical background spacetime  are the theoretical inputs to prediction in this paper. Sometimes we just denote the input theory by $(H,\Psi)$.
\end{subsection}

\begin{subsection}{ Histories}
\label{histories}
A time history (or just `history' for short) is  a sequence of alternatives at a series of times. Consider histories of the motion of the Earth's center of mass. A  coarse-grained history of this motion is specified by which of an exhaustive set of regions of space the center of mass occupies  at a discrete series of times. The  histories are {\it coarse-grained} because they only follow the center of mass of the Earth to an accuracy limited by the size of the regions  and   because they follow them,  not at all times, but only at a discrete series of them.  We next describe how such coarse-grained histories are represented in DHQM..

 An exhaustive set of coarse-grained yes/no alternatives at {\it one time} $t$ is
represented in the Heisenberg picture  by a set of orthogonal projection operators $\{P_\alpha(t)\}, \alpha=1,2,3\cdots$. 
%\begin{equation}
%\{P_\alpha(t)\} ,  \quad  \alpha = 1, 2, 3 \cdots  
%\label{projections}
%\end{equation}
These satisfy
\begin{equation}
\sum\nolimits_\alpha P_\alpha(t) = I, \  {\rm and}\quad \ P_\alpha(t)\, P_\beta (t) =
\delta_{\alpha\beta} P_\alpha (t) \ ,
\label{twoone}
\end{equation}

A one-dimensional $P_{\alpha}$  projects on a single pure state and is said to be {\it fine-grained}.  Projections $P_\alpha$ that project on bigger subspaces of $\cal H$ are {\ it coarse-grained} as in the discussion of the motion of the Earth's center of mass above. 

In the example of the Earth's center of mass motion  each of the projection operators in a set at time $t$ would project on a different spatial region at that time. For example the  projection $P_\beta(t)$   represents  the question ``Is the Earth's center of mass in region $\beta$ at time $t$ --- yes or no?  How time sequences of such  sets are turned into operators representing time histories of the Earth's center of mass motion is described immediately below. 

In the Heisenberg
picture, the operators $P_\alpha(t)$ evolve with time according to
\begin{equation}
P_\alpha(t) = e^{+iHt/\hbar} P_\alpha(0)\, e^{-iHt/\hbar}\, .
\label{twotwo}
\end{equation}
The state $|\Psi\rangle$  and the  Hamiltonian $H$ are  unchanging in time in the Heisenberg picture. .

An  operator $P_\alpha$ projects on some subspace of the Hilbert space. We  say that $P_\alpha$ {\it follows} the states in the subspace and {\it ignores} ones orthogonal to it. 

The ignored variables are important  because they can constitute an`environment' whose interactions with  followed variables carry away phases between followed histories and effect decoherence. For detailed examples of how this works see \cite{Ceq}, and (\cite{GH14}, Section VI.)

Simple examples of  sets of alternative histories can be specified by giving sets of single  time alternatives  $\{\{P^1_{\alpha_1} (t_1)\}$,
\{ \ $\{P^2_{\alpha [_2} (t_2)\}, \cdots$, $\{P^n_{\alpha_n} (t_n)\} \ \}$ at a sequence of times
$t_1<t_2<\cdots < t_n$.  The sets at distinct times can differ and are distinguished by the
superscript on the $P$'s. For instance, projections on ranges of position might be followed
by projections on ranges of momentum, etc.  An individual history $\alpha$ in such a set is a
particular sequence of alternatives $(\alpha_1, \alpha_2, \cdots, \alpha_n)\equiv \alpha$
and is represented by the corresponding chain of projections called a 
`class operator', viz.
\begin{equation}
C_\alpha\equiv P^n_{\alpha_n} (t_n) \cdots P^1_{\alpha_1} (t_1)\ , \quad \sum_{\alpha} C_{\alpha} =1.
\label{twothree}
\end{equation}
A history with at least one coarse-grained projection is said to be a coarse-grained history.
For any individual coarse-grained history $\alpha$, there is a {\it branch state vector}  defined by
\begin{equation}
|\Psi_\alpha\rangle \equiv  C_\alpha |\Psi\rangle, \  \quad \sum_{\alpha}|\Psi_{\alpha}\rangle =|\Psi\rangle.
\label{twofive}
\end{equation}
When probabilities $\{p(\alpha)\}$ can  assigned to the individual histories in a set,
they are given by
\begin{equation}
\label{probhist1}
p(\alpha) = \parallel |\Psi_\alpha\rangle\parallel^2 = 
\parallel C_\alpha |\Psi\rangle\parallel^2 = ||P^n_{\alpha_n} (t_n) \cdots P^1_{\alpha_1} (t_1)|\Psi\rangle||^2 .
\end{equation}

Negligible interference between the branches of a set 
\begin{equation}
\langle\Psi_\alpha|\Psi_\beta\rangle\approx 0\ ,  \ \  {\text  for\ all}  \quad  \alpha \not= \beta
 \quad(\text{\rm decoherence})
\label{twoseven}
\end{equation} 
%\begin{equation}
%\langle\Psi_\alpha|\Psi_\beta\rangle\approx 0\ ,   \quad  ({\rm all} \alpha) \not=\beta}.) 
% \quad(\text{\rm decoherence})
%\label{twoseven}
%\end{equation}
is a necessary and sufficient condition for the probabilities \eqref{probhist1} to be {\it consistent} with the
rules of probability theory \cite{Gri84}   --- whence the term `{\it consistent histories}'.   The orthogonality of the branches is approximate in
realistic situations. We mean by \eqref{twoseven} equality to an accuracy that defines probabilities well beyond the
standard to which the conditions  can be checked or modeled. A set of alternative  histories that satisfies the condition \eqref{twoseven} is said to {\it decohere} or be {\it decoherent}.

To sum up::  {\it Decoherence of a set of alternative histories is necessary for the probabilities of those histories to be  consistent with the usual rules of probability theory. Coarse raining is necessary for decoherence. } It is a remarkable fact that in quantum mechanics some information has to be ignored in order to have any information at all. 

{\bf   We call a decoherent set of alternative coarse-grained histories a {\it realm}}. 

It is instructive to rewrite the expression \eqref{probhist1} for the probabilities of a set of alternative histories in the Schr\"odinger picture.   We denote the Schr\"odinger picture representitive of a projection $P_{\alpha}(t)$ by ${\hat P}_{\alpha}$. Specifically,
\begin{equation}
{\hat P}_\alpha \equiv e^{iHt/\hbar} P_\alpha(t)\, e^{+iHt/\hbar}\, .
\label{schpict1} 
\end{equation}
%\begin{equation} 
%p(\alpha) = ||\hat\Psi_(\alpha_n\alpha_{n-1} \cdots \alpha_1\rangle||^2
%\end{equation}
Assume that the intervals $\Delta t$ between  time steps are equal in \eqref{probhist1}. Defining  $\tau\equiv\Delta t/\hbar$.  we can write for  \eqref{probhist1}. 
\begin{equation}
\label{sch2}
p(\alpha) = ||\hat P^n_{\alpha_n} (e^{-iH\tau}) \hat P^{n-1}_{\alpha_{n-1}} \cdots  \hat P^2_{\alpha_2} (e^{-iH\tau}) \hat P^1_{\alpha_1} (e^{-iH\tau})|\hat\Psi\rangle ||^2 .
\end{equation}

The expression for the history  on the right hand side of \eqref{sch2} could be described as unitary evolution represented by 
$e^{-iH\tau}$ interrupted by `reductions' represented by  the projections $\hat P_{\alpha}$'s  But these reductions have nothing to do with measurements in general. They are just the way  alternatives at one moment of time are represented in the   process of calculating probabilities for histories including histories of measurement situations e.g.(\cite{Wig63}, and Sections \ref{meassit} and \ref{measdsit}. The projections do not represent physical processes occurring in the measuremenrt interaction between apparatus and measured subsystem. Rather they are part of the description of of histories of measurement interactions whose probabilities are being calculated. 

Thus we recover the first of von Neumann's laws of evolution in a measurement situation but not the second --- state vector reduction on measurement, 
 
\end{subsection}

\begin{subsection} {Prediction and Retrodiction.}
In this subsection we make explicit  how DHQM can be used to make predictions of the future and retrodictions of the past. To do that let's fix a particular Lorentz frame in the assumed flat background spacetime. Suppose we are  at rest at the  the present moment $t_p$ with a certain amount of data about what happened in the past. The past is towards the big bang and  past moments are labeled by values of $t$  less than $t_p$  The future is the direction is away from  the big bang and moments there are denoted by values larger than $t_p$. (For more on what  past, present, and future  mean  in a closed quantum system  see \cite{now}). 

Suppose  that alternatives $\alpha_1$ and $\alpha_2$ have happened at times $t_1$ and $t_2$ in the past.  The probability that an  alternative $\alpha_3$ will happen at a future time $t_3$ given that alternatives $\alpha_1,\alpha_2$ have happened in the past is 
\begin{equation} 
\label{prediction} 
p(\alpha_3,t_3 |\alpha_2, t_2,\alpha_1,t_1)  = \frac{p(\alpha_3,t_3,\alpha_2,t_2,\alpha_1,t_1)}
{p(\alpha_2,t_2,\alpha_1,t_1 )} \quad  (\text{\rm prediction} )
\end{equation}
where all these probabilities are computed from $(H,|\Psi\rangle)$ via \eqref{probhist1}. 

Retrodiction is through similar formulae for conditional probabilities for past alternatives given present alternatives. For example suppose we know just that $\alpha_3$  happened at a present time $t_p=t_3$. The probability that $(\alpha_2, \alpha_1)$ happened at times $(t_2,t_1)$ in the past is.
 \begin{equation} 
\label{retrodiction} 
p(\alpha_2, t_2, \alpha_1,t_1 |\alpha_3,t_3 )  = \frac{p(\alpha_3t_3 \alpha_2,t_2,\alpha_1,t_1)} 
{p(\alpha_3,t_3)}\quad  (\text{\rm retrodiction} )
\end{equation}

\end{subsection}

\subsection{The Quantum State at a Moment of Time}
\label{qsmom}
        
Future predictions  can all be obtained from an effective 
density matrix summarizing information about what has happened by the present time.  For the example in \eqref{prediction}  suppose $t_2$ is the present time and alternatives $\alpha_1, \alpha_2$ are known to have happened at past times $t_1,t_2$ .we can  define an  effective density matrix $\rho_{\rm eff}(t_2) $ by 
%\begin{equation}
%\label{rhoeff1}
%\rho_{\rm eff} (t_2) \equiv    {P^2_{\alpha_2} (t_2)} {P^1_{\alpha_1} (t_1)} \rho %{P^1_{\alpha_1}(t_1)} {P^2_{\alpha_2}(t_2)}
%\end{equation}

%\begin{equation}
%\rho_{\rm eff} (t_2) \equiv  {P^2_{\alpha_2} (t_2)} {P^1_{\alpha_1} (t_1)} \rho {P^1_{\alpha_1}(t_1)} {P^2_{\alpha_2}(t_2)}
%\end{equation}

%\begin{equation}\frac{ {P^2_{\alpha_2} (t_2)} {P^1_{\alpha_1} (t_1)} \rho {P^1_{\alpha_1}(t_1)} {P^2_{\alpha_2}(t_2)}}{ {P^2_{\alpha_2} (t_2)} {P^1_{\alpha_1} (t_1)} \rho {P^1_{\alpha_1}(t_1)} {P^2_{\alpha_2}(t_2)}}

%\begin{equation}
%\rho_{\rm eff} (t_2) \equiv \frac{{P^2_{\alpha_2} (t_2)} {P^1_{\alpha_1} (t_1)} \rho {P^1_{\alpha_1}(t_1)} {(P^2_{\alpha_2}(t_2)}}{{P^2_{\alpha_2} (t_2)} {P^1_{\alpha_1} (t_1)} \rho {P^1_{\alpha_1}(t_1)} {P^2_{\alpha_2}(t_2)})}
%\end{equation}

\begin{equation}
\label{rhoeff}
\rho_{\rm eff} (t_2) \equiv \frac{{P^2_{\alpha_2} (t_2)} {P^1_{\alpha_1} (t_1)} \rho {P^1_{\alpha_1}(t_1)} {(P^2_{\alpha_2}(t_2)}}{{Tr[P^2_{\alpha_2} (t_2)} {P^1_{\alpha_1} (t_1)} \rho {P^1_{\alpha_1}(t_1)} {P^2_{\alpha_2}(t_2)})]}
\end{equation}

%\begin{equation}
%\rho_{\rm eff} (t_2) \equiv  \frac{{P^2_{\alpha_2} (t_2)} {P^1_{\alpha_1} (t_1)} \rho {P^1_{\alpha_1}(t_1)} {P^2_{\alpha_2}(t_2)}{D} 
%\end{equation}

 %{Tr(P^2_{\alpha_2} (t_2) P^1_{\alpha_1} (t_1) \rho P^1_{\alpha_1}(t_1)  P^2_{\alpha_2}(t_2))}
%\end{equation}
where $\rho\equiv |\Psi\rangle\langle\Psi |$.
Then  
\begin{equation} 
\label{pred}
p(\alpha_3,t_3 |\alpha_2, t_2,\alpha_1,t_1)  =Tr[ P^3_{\alpha_3}(t_3)\rho_{\rm eff}(t_3)]
\end{equation}
in agreement with \eqref{prediction}.

The density matrix $\rho_{\rm eff} (t_2)$ represents the usual notion of
state of the system 
at time $t_2$. It is given here in the Heisenberg picture and is constant  after $t_2$ until further information is acquired and  a new $\rho_{\rm eff}(t_3)$ must be
used for future prediction. See section \ref{collapse}. % Its Schr\"odinger picture representative which
%evolves with time, would be given by
%\begin{equation}
%\label{sch-evol}
%e^{-iH(t-t_k)/\hbar} \rho_{\rm eff}(t_k) e^{iH(t-t_k)/\hbar}, \quad.   t_k < t<t_{k+1}
%\end{equation} 

In contrast to prediction, there is no effective density matrix representing
present information from which probabilities for the past can be derived.  Probabilities for past history require knowledge of {\it both} present data {\it and} the quantum state
of the universe.

\subsection{The Quasiclassical Realm}
\label{qcrealm} 
Perhaps the most striking feature of our quantum universe is its  quasiclassical realm --- the wide range of time, place and scale on which the deterministic laws of classical physics apply to an excellent approximation \cite{qcrealm}. This includes the universe's classical spacetime geometry extending from just after the big bang to the distant future.  In DHQM the quasiclassical realm  can be derived from suitable theories of the universe's quantum state and dynamics $(H,\Psi)$. To derive classical spacetime the theory $(H, \Psi)$ must include quantum gravity in some approximation. \cite{qcrealm,Sau10}. 

So manifest is the quasiclassical realm that parts of it are routinely simply assumed in constructing effective physical theories that apply in the late universe. Classical spacetime, for instance, is the starting assumption for the standard model of the elementary particle interactions. Classical spacetime obeying the Einstein equation is assumed in cosmology to reconstruct the past history of our universe. Classical spacetime is assumed in CQM just to define the time $t$ in the Schr\"odinger equation, the times of   measurement, the time of state vector collapse, etc.  In Everett formulations classical spacetime is usually assumed to define the branching histories which are their characteristic feature. 

\subsection{Measurement Situations}
\label{meassit}
Whenever there is an alternative correlated with an alternative of our Universe's quasiclassical realm we say that we have a {\it measurement situation}.  From the value of the alternative of the quasiclassical realm we can infer the value of the correlated alternative.  In he Stern-Gerlach example the value of a spin (a quantum variable) becomes correlated with the spin's position in space (a classical variable.)  

Consider the example of a fission track in a slab of mica left by the decay of a radioactive nucleus embedded in the slab at its formation long ago. The track is produced by a quantum mechanical decay but  described in quasiclassical variables. There was `preparation' a long time ago, but not by human IGUSes. There has been irreversible amplification. The track is a record of what happened.  The decay and track is a measurement situation.

Correlation between two alternatives of the quasiclassical realm is a classical  measurement  situation. Correlation between an alternative of the quasiclassical realm and a `quantum variable' like a spin is a measurement situation whose results probe quantum theory.

Note that there is no requirement that anything as sophisticated as a human observer read the record. The track might have existed for milions of  years before having been noticed by a human being,  Measurement situations do not necessarily require anything as sophisticated as  human observers to be one element of the correlation.

\subsection{ Predicting Measurement Outcomes in DHQM}
\label{measdsit} 
DHQM is a comprehensive quantum theory that predicts probabilities for  suitably coarse-grained time histories of  what goes on in a closed system --- most generally the Universe. Measurement situations play no preferred role in the formulation of DHQM. %But DHQM can  predict the probabilities for  the outcomes of any particular measurement situation. 
%Measurement situations play no preferred role in the formulation of  DHQM.  
 But the probabilities of outcomes of measurement situations inside the closed system can be predicted in DHQM with decoherent sets of alternative histories that describe the interactions  between the subsystem being measured (`the system') and the subsystem doing the measuring (`the apparatus') so as to  produce  a record of the outcome accessible to relevant IGUSes including human observers.\footnote{  A more detailed example of this can be found in \cite{ Jer}  Sections II.9 and II.10.} Sequences of repeated measurements are described in DHQM by decoherent sets of histories that describe a series of such interactions and the creation of a record  record for each instance..    
Standard measurement models e.g. \cite{LB39,Roig}  show how this all works. 
\subsection{Who Ordered DHQM?}
\label{who-ordered}

After working through this brief exposition of DHQM a 
reader may well ask where  DHQM came from.  ``Who ordered that?'' 
The answer is that it is an inescapable inference from both the experimental and theoretical physics of the last century together with our large scale cosmological observations  that we live in quantum mechanical Universe. We therefore need a formulation of quantum mechanics that is adequate for cosmology.  That cannot just be a quantum mechanics of measurement situations. Specifically, we need  a generalization of CQM  with the following attributes:          
                          
* \ A  quantum mechanics of a closed system in which measurements and observers can be described but are not central to the formulation of the theory.

 *\  A formulation of quantum mechanics general enough to deal with quantum spacetime geometry in particular that near  the big bang, and what emerged from it.

*\ A formulation of quantum mechanics general enough to retrodict histories of the past of the Universe back to the big bang  to simplify the prediction of the future \cite{pasts}.

*\  A  formulation of quantum mechanics that can describe the emergence of classical spacetime in the early early Universe and, along with it  its quasiclassical realm. 
cf. \cite{GQMQC},  and with those both predict the future and retrodict the past. 

* \  A  formulation of quantum mechanics to which the Copenhagen quantum mechanics is an approximation appropriate for measurement situations.

 DHQM is such a formulation. DHQM is logically consistent, consistent with experiment as far as is known, applicable to cosmology, consistent with the rest of modern physics including special relativity, quantum field theory,  general relativity, and is generalizable to include quantum gravity (at least semiclassicaly) so that its capable of predicting the emergence of classical spacetime, and with that  both predict the future and retrodict the past.  It is unlikely to be  the only formulation with these properties  {cf. \cite{exprobs}),  but it is perhaps the most developed.

 \subsection{CQM from DHQM}
 \label{CQMder}
 Within DHQM of framework an approximate quantum mechanics of measurement situations,  aka CQM, can be derived. A measurement situation is a correlation between one alternative and an  alternative of the quasiclassical realm described in  Section \ref{qcrealm}..  If one is known the other is also.   Measurement models  show how this woerks.  An example  can be found \cite{Jer} Sections II.9 and II.10, and  \cite{Roig}.
 
 In models where a record of the outcome of the measurement is produced the probability of a history of the value of the record is all that is needed to recover a part of CQM. 
 
 But  there is also a part of CQM that does not follow from DHQM, namely the second law of evolution --- state vector reduction on completion of an `ideal' measurement. 
 As far as the author knows this is not inconsistent with theory or experiment. 
 
 CQM and DHQM are not separate independent  theories. 
CQM is best understood as an {\it approximation} to DHQM  for measurement situations. Therein lies the utility of DHQM to issues that arise in CQM that are the subject of this paper.

\subsection{A Note on Classical Spacetime}
\label{lass-st}
As mentioned above the classical spacetime of our quantum Universe is an essential component of the quasiclassical realm. Many of laws of classical physics follow from its local and global symmetries. The Navier-Stokes equations are an important example \cite{Ceq,GH06}. Another is the emergence of suitably coarse-grained classical spaceime obeying the Einstein  equation.

Classical behavior of anything is not a given in quantum mechanics. In a quantum universe classical behavior  is a matter of the quantum probabilities of appropriately defined histories supplied by the theories of the quantum state and dynamics $(\Psi,H)$. A quantum system behaves classically when these probabilities are high for suitably coarse-grained  histories of its evolution that exhibit correlations in time governed by deterministic classical laws. The classical behavior of the flight of a tennis ball, the orbit of the Earth, the collapse of a star to a black hole, and  the emergence and of  classical spacetime in the  early universe and its subsequent  evolution obeying the Einstein equation are all examples e.g. \cite{Har92,HHH08}.

\subsection{Final Theories}
\label{finth}
Many, if not most, of today's  proposals for `final theories' aim at a unification of all the forces of nature including gravity. To enable such a theory to predict our quasiclassical realm, including emergent  classical spacetime,  and also to retrodict what went  on in the very early universe where  classical spacetime breaks down,  requires a general enough formulation of quantum mechanics such as DHQM.  

Such a generalization of quantum mechanics is likely to be important for the experimental tests of such theories. The characteristic energy scale for unification is the Planck scale,
$\approx 10^{19} {\rm Gev}$. It seems unlikely that accelerators that reach this scale will be constructed on Earth in the near future. But these energies were reached in the big-bang. We can treat the big-bang as an experiment done once to test the predictions of these final theories with resulting data scattered over visible universe. 

\vskip .3in

\eject

\section{Buzzwords}
\label{bzwords}
This section treats the quantum buzzwords one by one in no very systematic order.
\vskip -.3in

\subsection{ States at  Moments of Time and The State of the Universe}
\label{st-mom}
 It is important to distinguish between a ``state at a moment of
time'' represented by $\rho_{\rm eff}(t)$ defined in  \eqref{rhoeff}
and the quantum state of the closed system (like the universe) represented by the Heisenberg $\rho=|\Psi\rangle\langle\Psi|$  for appropriate $|\Psi\rangle$.  A  theory of the quantum state $|\Psi\rangle$  is an essential input to any quantum mechanical theory describing a closed system. 
 A notion of ``state at a moment of time'' is a not a necessary part of that theory. 
DHQM  may be organized fully four-dimensionally in terms of histories without a notion of state at a moment of time. 
Indeed,  in quantum theories of spacetime, where spacetime geometry is fluctuating and without definite value near the big-bang   there there will be no ``state at a moment of time'' not least because there will not be a classical notion of `moment of time  near the  big-bang. Without classical spacetime  there will  be no CQM. Rather, we expect CQM to emerge  along with  a classical spacetime  in the evolution of the Universe to give meaning to  CQM's  notions of time,  and state at a moment of time \cite{Har92, Har94}. 

\subsection{IGUSES (Observers)  }
\label{obs}

In this paper, instead of the term `observer', we will use the more general term `IGUS'   (Information Gathering and Utilizing System).   This  emphasizes that whatever role such systems play in quantum mechanics they do not have to be human beings. Humans, both individually and in various collections, are examples of IGUSes.   There are many other examples --- self driving cars., certain computers, erc. A thermostat is a trivial  example. The collection of humans working in science we call the `human scientific `IGUS' or HSI for short. All IGUSes are physical systems within the Universe not outside it some sense. 

As physical systems within the universe, IGUSes  are subject to the laws of quantum theory.  But IGUSes are not a necessary part of formulating quantum theory. In DHQM for example, 
probabilities are predicted  alternative positions of the Moon whether or nor there is an IGUS looking at it or measuring it.     Quantum probabilities from the theory $(H,\Psi)$  predict  probabilities for  alternative values of density fluctuations in the very early universe whether or not they were participants in a measurement situation and certainly whether or not there was an observer registering their values.

\subsection{ State  Reduction}
\label{collapse}
As an IGUS  acquires new  information  the probabilities for prediction of the future like \eqref{prediction}  and the effective density matrix representing current information have to be updated.  Consider  the example of prediction represented by  the density matrix \eqref{rhoeff}.  Suppose by measurement or otherwise we find out that alternative $\alpha_3$   has happened at time $t_3$ and want to predict the probabilities of further events at later times.
The density matrix  $\rho_{\rm eff}(t_2)$  at time $t_2$ would be updated to time $t_3$ as follows:
 \begin{equation}
\label{update}
 \rho_{\rm eff} (t_3) \equiv  ({1{/\cN)}} {P^3_{\alpha_3} (t_3)}  { P^2_{\alpha_2}(t_2)}{P^1_{\alpha_1}(t_1)} \rho {P^1_{\alpha_1}(t_1)P^2_{\alpha_2}(t_2)}P^3_{\alpha_3}(t_3)
\end{equation}
where $\rho\equiv |\Psi\rangle\langle\Psi |$ and $\cN$ is the normalizing factor:
\begin{equation}
{\cN}\equiv Tr [ P^3_{\alpha_3} (t_3){ P^2_{\alpha_2}(t_2)}{P^1_{\alpha_1}(t_1)} \rho  {P^1{\alpha_1}(t_1)}{ P^2_{\alpha_2}(t_2)}P^3_{\alpha_3} (t_3) ] .
\end{equation}
This updating has been describe in various other ways. 
We could say that the sequence of projections in \eqref{update} has `reduced' the state $\rho$ or that $\rho$ has `collapsed' under their action, or that the state is following the second law of evolution.  It's the same meaning for all of these.

%\begin{equation}
%\frac{T}{B}\end{equation}

%P_{\alpha_1}(t_1) P_{\alpha_2}(t_2)} {Tr(P^2_{\alpha_2} (t_2) P^1_{\alpha_1} (t_1) \rho P_{\alpha_1}(t_1) P_{\alpha_2}(t_2))} 
%\frac{T}{B}      %{P^3_{\alph_3},t_3)}{B} 

%P^2_{\alpha_2} (t_2)}{B}{ P^1_{\alpha_1} (t_1)}{B} \rho P_{\alpha_1}(t_1) P_{\alpha_2}(t_2)} {Tr(P^2_{\alpha_2} (t_2) P^1_{\alpha_1} (t_1) \rho P_{\alpha_1}(t_1) P_{\alpha_2}(t_2))} 
%\end{equation}

%\begin{equation}
%\label{update}
%\rho_{\rm eff} (t_3) \equiv \frac{P^3_{\alpha_3}(t_3)  P^2_{\alpha_2} (t_2) P^1_{\alpha_1} (t_1)\rho }{B}

%\begin{equation}
%\label{upddate} 
%\rho_{\rm eff}(t_3)\equiv \frac{\alpha_3,t_3\alpha_2, t_2,\alpha_1,t_1)  \frac{p(\alpha_4,t_4, \alpha_3,t_3,\alpha_2,t_2,\alpha_1,t_1)}
%{p(\alpha_4,t_4, \alpha_3,t_3, \alpha_2,t_2,\alpha_1,t_1)}  %(\text{\rm prediction} )
%\end{equation}

Much has been made of this updating of the probabilities for prediction. We could say that the state $\rho=|\Psi\rangle\langle \Psi|$ has been ``reduced'', or ``collapsed'' by action of these projections on $\rho$. However, there is nothing specifically quantum mechanical about 
it.  It occurs in any statistical theory.  In a sequence of
horse races the joint probabilities for a sequence of eight races is naturally
converted, after the winners of the first three are known, into conditional
probabilities for the outcomes of the remaining five races by exactly this
process.  

In the quantum mechanics of a closed system,  this
second law of evolution''  or collapse of the state  $\rho_{\rm eff}(t)$ has no special status in DHQM  and no particular association with a measurement situation or any physical process.  It is simply a convenient way of organizing the time sequence of probabilities that are of interest to a particular IGUS.  Indeed,
DHQM can be formulated without ever mentioning ``measurement'',
``an effective density matrix'', its``reduction'' or its ``evolution''.  Further,   as in quantum theories of spacetime (quantum gravity),  there can  be situations
where it is not possible to introduce an effective density matrix at all, much
less discuss its ``evolution'' or ``reduction''. 

%It is in the ideal measurement model of Section II.10, upon which
%the Copenhagen approximation to quantum mechanics is based, that we can connect
%the reassessment of probabilities with the ``reduction of the wave packet'' on
%{\sl measurement}.  There, as can be seen from (II.10.8), [cf.
%(II.3.4)-(II.3.5)]  the effective density matrix.
%$$\rho_{s,{\rm eff}}(t_k) = \frac{s^k_{\alpha_k}(t_k) \cdots s^1_{\alpha_1}(t_1)
%\rho_s s^1_{\alpha_1}(t_1) \cdots s^k_{\alpha_k}(t_k)}{tr\left[s^k_{\alpha_k}
%(t_k) \cdots s^1_{\alpha_1}(t_1) \rho_s s^1_{\alpha_1}(t_1) \cdots
%s^k_{\alpha_k}(t_k)\right]}\eqno({\rm A}.1)$$
%summarizes present information for future prediction of the subsystem under
%observation.  {\sl Every} projection operator in (A.1) is part of a
%measurement situation in this idealized model.  That is, in the larger universe
%of apparatus and subsystem each projection  is exactly correlated with an 
%exactly
%decohering record variable.  Thus, it is possible to say that $\rho_{s,{\rm
%eff}}(t)$ is 
%constant in between measurements (in this Heisenberg picture), but is
`%`reduced'' {\sl at a measurement}.  
%Two remarks may be useful concerning the ``state vector reduction'' in
%the CQM.  First, again, the quantum mechanics of a 
%subsystem
%under observation may be formulated directly in DHQM in terms of probabilities for
%histories without an effective density matrix or its reduction.  
%Second, and more importantly, the association of the ``reduction'' with
%``measurement'' is a special property of the ideal measurement
%model.  This has suggested to some that there is a physical mechanism behind
%the reduction of the wave packet.  However, in the more general situations
%in which a closed system is considered, there is no necessary association of
%``reduction'' with a measurement situation.
%\subsection{Does Everett Eliminate Reduction}
%\label{everett-red}
   
\subsection{Consciousness}
\label{conc}
Does consciousness play a role in formulating a quantum mechanics of measurement situations? Some distinguished theorists have said something like this\footnote{  E. Wigner, J.A. Wheeler, D. Page, H. Stapp, and R. Penrose and even the author early on  in \cite{ Har68} come to mind.}.  CQM is  a theory of a single subsystem whose state evolves by von Neumann's two laws of evolution. --- unitary evolution when it is isolated and state vector reduction when it is (ideally) measured\footnote{In his book unitary evolution was the second of the two laws \cite{vonN} and reduction the first, but modern practice has turned these around}. These processes can alternate in a history of successive measurements situations.  But at least some of the founders of the subject thought that the series ended in a final ``reduction in consciousness'' {e.g.  \cite{Wig63,Wigpc}.   This is problematical since, as far as the author knows, `consciousness'  is a complex phenomenon that is imperfectly understood e.g. \cite{Gaz}. Consciousness plays no role in formulating DHQM. But nothing excludes DHQM from helping to understand consciousness,

In DHQM realistic measurement situations can be described in terms of histories of an apparatus that interacts with the measured subsystem and creates a classical record of the outcome of the measurement that is accessible to an IGUS. 
A measurement model illustrating this in the context of DHQM is described in Sections II.9 and II.10 of \cite{Jer}. Consciousness is not involved.

\subsection{Does `Measurement'  Need  a Mathematically Precise Definition?}
\label{defmeas}
Measurement plays a fundamental role in the formulation of CQM. It is therefore natural to ask that  `measurement' be precisely and 
mathematically defined. Essential features of measurement have been seen to be:  irreversible amplification beyond a certain level, association with a
macroscopic variable, a further association with a long chain
of such variables, and the formation of enduring records.  Efforts have been
made to attach some degree of precision to words like ``irreversible'',
``macroscopic'', and ``record'', and to discuss what level of ``amplification''
needs to be achieved  and how much the entropy has to go up. for irreversibility.  The author's impression is that no precise  definition has been universally accepted. But the author also knows of no mistake that has been made by this absence of a mathematically precise definition over the more than a century since CQM was first formulated.

\subsection{CQM As an Approximation to DHQM}
\label{recovery} 
Measurement  is not a fundamental notion in  the formulation DHQM so a precise mathematical definition is neither required nor possible.  The probabilities  of the outcomes of specific measurement situations can be predicted in DHQM as described in Section \ref{measdsit}.  A more detailed example  can be found in \cite{ Jer}  Sections II.9 and II.10.}  Sequences of  repeated measurements are described in DHQM by sets of histories that describe a series of such interactions.  Standard measurement models e.g. \cite{LB39,Roig}  suggest  how this all works. Measurements and the CQM that describes them are approximations to DHQM valid in particular  circumstances.
Once measurement is seen as an approximate notion there is less of an imperative for its precise mathematical characterization.
What is {\it not}   recovered from DHQM is the second law of evolution --- state vector reduction. As far as the author knows this is not in conflict with experiment or observation.

\subsection{Many Worlds} 
\label{manyworlds}
DHQM predicts probabilities for the individual members of decoherent sets of coarse-grained 
alternative} histories of the universe.  Within a given set one cannot assign
``reality'' simultaneously to different alternatives because they are
contradictory.  Everett \cite{Eve57}, DeWitt  \cite{DeWitt70}, and others have described this
situation, not incorrectly, but in a way that has confused some, by saying that
all the alternative histories are ``equally real''.  What is meant is that
quantum mechanics prefers no alternative history over another except through its
probability \cite{Wal12}. 

Probabilities may be assigned to alternative positions of the Moon and to alternative values of density fluctuations whether or not they were participants in a measurement situation and certainly whether or not there was an observer registering their values.
 
 \subsection{CQM vs Everett?}
 \label{CvE}
 CQM is not in any sense opposed to Everett's idea of applying quantum mechanics to closed systems like the Universe. It is an approximation to the more general framework  of DHQM which {\it is} consistent with Everett's ideas. It is an approximation that is appropriate in the special cases of measurement situations in which the observer is treated as a physical system within the Universe.
 From this point of view the many successes CQM should be seen as supporting DHQM,
 and along with it, the Everett viewpoint. 
 
 \subsection{The Origin of the Quasiclassical Realm}
 \label{orig-qcrealm}
 Suppose our  Universe is understood  fundamentally by a unified quantum theory of dynamics in $H$ and a quantum theory of its state $|\Psi\rangle$  both characterized by the Planck scale.  What is the origin of our universe's  quasiclassical realm?  Some key steps in its emergence are the following. \cite{qcrealm,Ceq, harchain} :
 \begin{itemize}
 \item{} Classical spacetime emerges from the quantum fog at the beginning.
 
 \item{} Local Lorentz invariance of the classical spacetime implies conservation laws for energy, momentum, number, etc for fields moving in the spacetime.
 
 \item{} Sets of alternative histories of averages of densities of conserved quantities  over  suitably small volumes (quasiclassical variables)  decohere.  .

 \item{} Approximate conservation implies that quasiclassical variables are predictable despite the noise from mechanisms of decoherence. 
 
 \item{}  Local equilibrium implies closed sets of classical equations of motion \cite{GH06,Hal98,Hal99}. 

 \end{itemize}
 
 \subsection{When Does CQM Emerge  in the History of the of the Universe?}
 \label{when-cop}
 CQM is most naturally applicable to laboratory measurement situations. It relies on  a number of features of our Universe notably its quasiclassical realm  described in  Section \ref{qcrealm},  that includes  the classical spacetime in which IGUSes evolve,  operate, build laboratories, carry out experiments,  and otherwise act  like  observers. These prerequisites are not available in all epochs of the Universe's history.  They emerge over time as described in Section \ref{orig-qcrealm}
 We can therefore pose the question {\it when} in the history of our Universe  did  CQM emerge as an approximation to DHQM \cite{Har16}. 
 
 To answer crudely divide cosmological time roughly up into four epochs:
 
 \begin{itemize}

\item{\it The Epoch of Quantum Gravity:}  A short era $t\lesssim 10^{-43} s$ in which the geometry of spacetime and matter fields exhibited large quantum fluctuations.  The histories constituting a quasiclassical realm including  classical spacetime   start only after this period.

\item{\it The Early Universe:}  The early universe contained  a hot plasma of nucleons, electrons, neutrinos, and photons. As revealed by the observations of the cosmic background radiation (CMB) in this epoch the universe was homogeneous, isotropic and featureless to an excellent approximation (deviations from exact isotropy of $1$ pt in $10^5$).  

\item{\it The Middle Universe:} As the universe expands the temperature drops. When it has dropped sufficiently,  electrons, protons and nucleons recombine and nuclei are synthesized. The initial  fluctuations from the quantum gravity era grow under the action of gravitational attraction. Eventually they collapse under the forces of gravitational attraction to produce the universe of galaxies, stars, planets, biota and IGUses that we find today.uared

\item{\it The Late Universe:}  The cosmological constant causes the universe to expand and cool exponentially quickly. Stars exhaust their thermonuclear fuel and die out. Black holes have evaporate. The  density of matter and  the temperature  approach zero. The universe is cold, dark and inhospitable.  

\end{itemize}

When in this history would one expect to find IGUSes, a quasiclassical realm, laboratories, experiments and CQM? Not in the quantum gravity epoch where there isn't even classical spacetime to define a notion of localized subsystem. Not in the early universe.  That period has regularity in its homogeneity and isotropy, but it lacks the complexity that would give an evolutionary advantage to being an IGUS \cite{Har16}. Not in the late universe for the same reason (and also  because of the failure environmental mechanisms of decoherence in the far future \cite{HH20}).  Only in the middle universe do we have both regularities for an IGUSes  to exploit and enough complexity to make exploiting those regularities fruitful. That is where we do find them (for more on this see \cite{Har16}). Thus only in the middle Universe would we expect CQM to be implementable  as an approximation to DHQM.

\subsection{The Principle of Superposition}
\label{superpos}
The principle of superposition is a fundamental part of any quantum mechanical framework.
It is incorporated into DHQM and its approximation CQM. 

Suppose, as in the two-slit example, there are two histories $A$ and $B$ with branch state vectors $|\psi_A\rangle$ and $|\psi_B\rangle$ leading to the same outcome,  cf  Section \ref{histories}. The probability of the outcome is the absolute  square of  sum of the branch state vectors. It is not the sum of the absolute squares of the branch state vectors unless the two vectors are orthogonal so the set of two decoheres.  The interference between the two branches results in characteristic interference patterns as in the two-slit case.\footnote{For much more detail by the author on the two-slit example see \cite{Har19}.}.

\subsection{ Is CQM  Non-Local?} 
\label{nonlocal}
 Situations such as that in the
  Einstein-Podolsky-Rosen  thought experiment
have suggested to some to that CQM is non-local. 
  However, it is straightforward to show very generally using
techniques of the present formulation that it involves no non-locality in the
sense of quantum field theory and no signaling outside the light cone. (For
alternative demonstrations cf. e.g.  \cite{GRW,Jor83}) and Section 5 of the Appendix `Buzzwords'  of \cite{Jer}).

Bohmian mechanics is an alternative formulation of quantum theory that is explicitly non-local but is said to be consistent with relativistic causality \cite{BH93}.   In at least one of its formulations it makes predictions which are different from DHQM e.g.\cite{Har04}.

\subsection{When Does a Subsystem Have a  Pure Quantum State?} 
\label{sub-pure?}
By `subsystem' we usually mean very roughly an approximately localized collection of matter interacting weakly with the rest of he Universe over a period of time. Stars, planets, trees, animals, bacteria are familiar examples. In classical physics a subsystem is described by classical variables which we can take to be averages of densities of energy, momentum, and number over the individual members of a partition of space into suitably small volumes. These are  called the `quasiclassical variables' or sometimes `hydrodynamic variables' like those  that occur in the Navier-Stokes equation for instance.

The question naturally arises as to whether such a subsystem is described by a quantum state in the Hilbert space of the whole. In very carefully prepared situations such as a spin  moving through a Stern-Gerlach apparatus the answer is `yes'.   But generally the answer is `no' . Subsystems generally are interacting with their environment and entangled with it.  After an an  electron in the two-slit experiment is detected by making  a mark on photographic plate it is entangled with other atoms in the screen where it is detected.

Indeed, if the interaction with an environment is the typical mechanism for  decohering sets of histories of the subsystem (for example coarse-grained  histories of its position ) it will always be entangled with the environment and not pure.

When does a subsystem have a quantum state?  Only in carefully designed laboratory situations. Almost never in generic realistic situations. 

\subsection{The Quantum Arrow of Time}
\label{qarrow}

In the introduction to DHQM in Section \ref{back} the expressions for the probabilities of histories like \eqref{probhist1} and \eqref{schpict1} are not time neutral.  The state $|\Psi\rangle$ on one end of the chain of projections and nothing on the other end.
That asymmetry is called the quantum mechanical arrow of time. It is built in to the the version of DHQM  described in Section \ref{back}. 

However, as noted by several authors e.g. \cite{ABL64,Har96,Har04, Har13,HH12}, DHQM can be formulated time neutrally and more generally be employing both initial and final density matrices that enter by symmetrically into the expressions like \eqref{probhist1}  for the probabilities of histories.  Then all of the arrows of time exhibited by the universe emerge from differences between these initial and final conditions even with  time neutral dynamics $H$.  These include the thermodynamic arrow (increasing entropy),  electromagnetic arrow (retarded radiation), the psychological arrow (we remember the past, experience the present and predict the future), the expansion of the universe arrow, the growth of  cosmological fluctuations arrow,  and the quantum arrow under discussion

Following vonNeuman \cite{vonN} CQM  is usually  formulated in terms of two laws. First, 
the law for the unitary evolution of the state vector described by the Schr\"odinger equation. Second, the law for the reduction of the state vector after a  measurement that disturbs the measured subsystem as little as possible. The Schr\"odinger equation can be run both forward and backward in time. However, state vector reduction can be run in only one direction in time --- usually assumed to be towards the future or  or in the direction of increasing entropy.   The second law of evolution  thus  singles out a direction in time defining the  quantum arrow of time.  See, e.g. \cite{ABL64,Har13,Har96}. 

DHQM of the universe can be formulated time neutrally with  initial and final conditions playing symmetric roles in the formulation \cite{Har96,Har13} and no built in arrows of time. Rather all arrows of time, including the quantum mechanical one,  are emergent features of the differences between the initial and final conditions of our particular Universe  e.g. \cite{Har13}

\subsection{Is  Schr\"odinger's Cat Paradoxical?}
\label{scat-paradox}
In the Schr\"odinger's cat thought experiment \cite{scat} the cat is a subsystem of the `universe' defined by the  containing box and its contents.  But the cat  is not an isolated subsystem  inside that box.. The cat  is interacting with particles of the walls of the box, with the electromagnetic radiation in the box, and  with the cyanide gas that is released when geiger counter clicks.  The cat is entangled with all those other things. We can think of an initial state where  the box is full of air, where  the geiger counter is turned on, and the cat is standing, alive, and another state where geiger counter  has clicked and the cat is dead on the floor of the box. These are exclusive alternatives represented by a set of  two orthogonal projection operators as in \eqref{sch2}  referring to the whole box plus cat which we can compactly  label ($P_{alive},P_{dead})$ adding to unity. There is no sense in which the cat is simultaneously both alive and dead.  That would be represented by the product of the two projections but that is zero. The state of the box is in a superposition of the histories in which the cat is dead and the cat remains alive. It's either or  ---  One of these happens and the other does not. There is no paradox.

In quantum cosmology a  state of the universe like the no boundary wave function \cite{NBWF}, is a superposition of different histories of what happens.  We (the HSI) are living in this  superposition.  We are all Schr\"odinger cats in the no-boundary quantum state of the universe. For more see Section \ref{livsup} just below. 

\subsection{Living in a Superposition}
\label{livsup}
A generic wave function of the universe will not predict a single history with unit probability. Rather it predicts decoherent {\it sets} of alternative histories  together with the probabilities of which history occurs. The wave function of the universe $\Psi$ is a superposition of  these \eqref{twofive}.  Physical systems within the universe like us can occur in one or possibly more of these histories.  We have no way of detecting or feeling the other histories  --- they are exclusive. We don't see the other things `smeared out' because everything we look at is in the histories of the ensemble along  with us.  cf. \cite{Sau10}. 

Occasionally even  distinguished scientists ask: `'If I am in la superposition of states why don't I feel that I am in a superposition?'' An interference pattern is an experimental signature of a superposition as in the two-slit experiment.  But we con't naturally come equipped with an interferometer that would enable either us or Schr\"odinger's cat 
to `feel' a superposition. for example.There isn't ny physical mechanism by which us or Schr\"odinger's cat could feel a superposition.  Indeed, from the perspective of quantum cosmology we are all Schr\"odinger cats  superposed in the wave function of the universe e.g \cite{NBWF}. A thought experiment where a human IGUS is  a participant in a huge two-slit experiment is described in \cite{superpos}.  Such an  IGUS could detect whether its state was a superposition by observing quantum interference. We could say with the late Bryce DeWitt: ``People who say that they don't feel themselves to be in a superposition are like the people who said that they didn't feel the Earth move in its orbit around the Sun''. 

\subsection{Human IGUSes and the Quasiclassical Realm}
\label{focus}
%\cite{qcrealm}

In DHQM  the  {\it quasiclassical realm} described in Section \ref{qcrealm}  is  a decoherent set of alternative histories  coarse-grained by values of quasiclassical variables such as the `hydrodynamic variables' briefly defined in Section \ref{sub-pure?}. The quasiclassical realm  includes  classical spacetime\footnote{ A quantitive measure of a realm's classicality was defined in \cite{measclass}. Roughly the. measure  is  the augmented entropy e.g. \cite{Zur89}  --- the sum of the entropy of the set  and the  algorithmic information content of its coarse-grained description.}. 

The quasiclassical realm of our universe exhibits a high level of predictability because of its  deterministic regularities. It  exhibits a high level of simplicity because its  histories are defined by a small set of `classical' variables operating on scales that are very small compared those of the large scales characterizing our  Universe. 
Both individually and collectively  human IGUSes, and the HSI in particular, are described in terms of quasiclassical variables. We are therefore not separate from our Universe's quasiclassical realm but rather part of it.

As human IGUSes we utilize almost exclusively the variables that define the quasiclassical realm, operate by its quasiclassical laws, and focus on features and subsystems of the Universe that can be described in quasiclassical terms \cite{qcrealm}.
Essentially all the data that we have about the universe are recorded in the variables of the quasiclassical realm. A theory is more successful when it predicts correlations in our data. For more discussion of how the correlations are predicted and the distinction between `top-down' and `bottom up' approaches to prediction see \cite{TDBU}.

Plausibly, both individually and collectively,  we and other living things evolved to exploit the  regularities that characterize the quasiclassical realm in order to satisfy the age-old imperatives:  get food --- yes, be food --- no, make more --- yes. That evolution is in principle described by sets histories that follow  biological evolution. These  are characterized by frozen accidents  such as mutations, recombination, and genetic drift. --- chance events with long term, widespread consequences.  The probabilities of most of these histories  are well beyond our power to compute or  measure today but see e.g \cite{Lenski03,WK00} and the remarks in \cite{qcrealm, Har16,Har04}.  

\subsection{Other Quasiclassical Realms, Other IGUSes?}
\label{otherrealms} 
A fascinating question is whether the Universe exhibits  realms coarse-grained by variables different from the usual quasiclassical ones that also have high levels of predictivity and simplicity  as judged by a quantitative measure like that for  classicality as discussed in \cite{measclass}.  Would such a realm have evolved  IGUSes that are different from the ones we find in the usual quasiclassical realm, and if so could we communicate with them?

\subsection{Separate Quantum and Classical Worlds? The Heisenberg Cut?}
\label{worlds}
Older formulations of CQM   sometimes assumed 
 a classical world and a separate quantum world, with a movable boundary  between the two called the Heisenberg cut e.g. \cite{LL58})  Observers and their measuring apparatus make use of the classical world, so that  the results of  a ``measurement'' are  ultimately expressed in one or more ``c-numbers''. 

In DHQM there is no such division and no such boundary.  All things are quantum but some things behave classically. A subsystem behaves classically when, as  consequence of (H,$\Psi$ ), it is described by  a suitably coarse-grained decoherent set of alternative histories of the universe whose probabilities favor correlations in time following classical deterministic laws. 

There is no Heisenberg cut. Classical behavior in our Universe is not something to be {\it posited}.  It is something to be {\it calculated} from $(H,\Psi)$. A separate classical world is neither needed nor correct --- it  it  is all quantum mechanical. Classical behavior emerges from quantum mechanics and $(H,\Psi)$--- it's all quantum mechanical. 

\subsection{Beyond Measurement Situations  -- Quantum Cosmology}
\label{beyondmeas}
Its an inescapable inference from the physics of the last century that we live in a quantum mechanical universe. If so, there is no escaping generalizing CQM to provide a formulation  general enough to be  applicable to cosmology. Too much our understanding of the universe depends on such a generalization --- the quantum nature of the big bang, the origin of coarse-grained classical spacetime, the origin of the large scale structure seen in the CMB and the distribution of galaxies  in early quantum the existence of  localized systems  including IGUSes,  etc. and, indeed, the origin of, IGUSES, and  CQM itself. 

DHQM is such a generalization \cite{Jer,leshouches}. DHQM is logically consistent, consistent with experiment as far as is known, consistent with the rest of modern physics such as special relativity, and quantum field theory, general enough or cosmology, and generalizable to apply to semiclassical quantum spacetime  \cite{leshouches}.  It can both predict the future and retrodict the past of our Universe.  It is  not  the only quantum framework for cosmology (e.g \cite{LPVP} )but it is one of the the most developed of the  ones we have at present.

CQM, deals with  measurement situations.  Predictions of CQM are therefore predictions of DHQM. But beyond measurement situations DHQM is the basis for predictions in cosmology of our large scale observations of the universe from quantum theories of that follow from the theory $(H,\Psi)$  {e.g. quantum field theory.)   These predict  for example,  the universe's classical spacetime, its approximate homogeneity and isotropy on scales above several hundred Mpc today, and  deviations from these symmetries that we see today  in the fluctuations of the cosmic background radiation  and  in the large scale distribution of the galaxies.  See e.g. \cite{HHH08}.

\subsection{Final Theories}
\label{finth}
Many, if not most, of today's candidates for final theories aim at a unification of all the forces of nature including gravity.  The Planck  energy is their characteristic energy scale. Such  a theory must predict our observable quasiclassical realm including its classical spacetime and also to retrodict  what goes on in the very early universe where  classical spacetime breaks down. The early universe may be the only place where Planck scale energies that are characteristic of these theories to occur in our Universe.  One can therefore think of the big bang as an experiment in which Planck scale energies are reached with data on the outcome spread over a large part of universe now  that can be used to test such theories. Quantum cosmology seems destined to play an important role in in the physics of the elementary particles.

\section{Conclusion}
\label{conc}

\subsection{No Escape from Generalizing CQM \\ for Cosmology, Final Theories, and Unification}
\label{noescape}

As discussed in the Introduction and  in \cite{Har19} there is   is no escaping generalizing  CQM to a quantum framework applicable to cosmology and to quantum spacetime.  Too much of our understanding of the Universe depends on it  --- the quantum nature of the big-bang, the quantum origin of classical spacetime, the origin of large scale structure in the  gravitational collapse of early quantum fluctuations, the emergence  of  localized systems, and  the origin of IGUses through the frozen accidents of biological evolution. More succinctly,  we need a generalization  of CQM to explain the emergence of the quasiclassical realm of everyday experience including emergent  classical spacetime and,  indeed, to explain the emergence of  the CQM approximation itself.

Further, a generalization  of CQM is  needed to calculate  and test the predictions of final theories that unify the quantum theories  of fundamental interactions with quantum gravity. In short we need a generalization of CQM to unify the physics of the very large with the physics of the very small. DHQM is one such generalization but probably not the only one.

This paper briefly explored  the idea that many of the difficulties with understanding the CQM of measurement situations can be clarified by examining its generalization --- DHQM.  The  aim has been has been to show by many short expositions a certain unity in the explanations and to briefly suggest starting points for further research. 

The utility of generalizations for improving understanding is not surprising. Similar clarifications have been instructive  elsewhere in physics. We understand thermodynamics better with the help of statistical mechanics. The  solar system regularities of Ptolemaic astronomy were better understood from Newtonian mechanics and the gravitational force law.   The nature of gravity and of the spacetime in which we live are best understood  from Einstein's theory of gravity --- general relativity --- and the distribution of  mass-energy in the Universe. There are many other examples. These understandings are confirmed by the new observations that they successfully predict  --- gravitational waves are a recent example. 

\subsection{Beyond DHQM?}
\label{beyondDH}

Despite its successes some find DHQM  unsatisfactory
by  standards for physical theory beyond logical consistency and  consistency with experiment and observation.
The intuition of others suggests that in domains where the predictions of quantum mechanics have not yet been fully tested an experimental inconsistency will emerge and a different theory will be needed.  For example, perhaps the interference between
``macroscopically'' different configurations predicted by quantum mechanics
will not be observed on large distance scales. e.g. \cite{Pen14}.  And indeed the application of DHQM to cosmology is an enormous  and mostly untested extrapolation. What is needed
to meet such standards, or to resolve such experimental inconsistencies should they develop, is not further research on DHQM  itself, but rather  new and
conceptually different theoretical frameworks.  It would be of great interest to
have serious and compelling alternatives to DHQM   e.g \cite{LPVP} if only to suggest decisive
experimental tests both of DHQM and other formulations of quantum mechanics.  We need more research both theoretically and experimentally!

\subsection{The Relation Between DHQM and CQM} 
\label{DCHconnect}
Did we derive CQM from DHQM in this paper?  We did not. Even assuming classical spacetime that would require having a general analysis of measurement situations and the formation of accessible records of their outcomes not to mention the quantitative analysis of measurement models. 

We did suggest in Section \ref{CQMder}, especially Section \ref{recovery}, how measurement situations could be analyzed in DHQM using measurement models based on decoherent sets of histories describing the interaction between a measurement apparatus and a measured subsystem.  But we found no evidence that  DHQM supports a general second law of state vector reduction.\footnote{And indeed the author has not been able to identify decisive experimental evidence for a general second law applicable to all measurement situations  outside simple systems like the Stern-Gerlach model.} 

Progress can be made by examining more and different measurement models such as those in \cite{Jer} (Sections 9 and 10) and \cite{Roig}. There is still much to do.  

\subsection{Concluding Manifesto}
\label{manifesto}

The author ends this work by quoting the concluding paragraph of his first paper on DHQM with Murray Gell-Mann \cite{GH90}:
``We conclude that resolution of the problems of interpretation presented by quantum mechanics is not to be accomplished by further intense scrutiny of the subject as it applies to reproducible laboratory situations, but rather through an examination of the origin of the Universe and its subsequent history. Quantum mechanics  is best and most fundamentally understood in the context of quantum cosmology.''    

\section{Appendix A:  Copenhagen Quantum Mechanics and Decoherent Histories Quantum Mechanics. }

{\it Another perspective on the author's view on the relationship between CQM and DHQM is provided by the following modestly edited excerpt from the conclusion of a paper with Murray Gell-Mann:  `` Quasiclassical Coarse Graining and Thermodynamic Entropy'' } \cite{GH06}

 This appendix is concerned with the relation between our approach to quantum mechanics, based on coarse-grained decoherent histories of a closed system, and the approximate quantum mechanics of measured subsystems, as in the ``Copenhagen interpretation.''
The latter formulation {\it postulates} (implicitly for most authors or explicitly in the case of Landau and Lifshitz \cite{LL58}) a classical world and a quantum world, with a movable boundary  between the two. Observers and their measuring apparatus make use of the classical world, so that  the results of  a ``measurement'' are  ultimately expressed in one or more ``c-numbers''. 

We have emphasized that this widely taught interpretation, although successful, cannot be the fundamental one because it seems to require a physicist outside the system making measurements (often repeated ones) of it. That would seem to rule out any application to the universe, so that quantum cosmology would be excluded. Also billions of years went by with no physicist in the offing. Are we to believe that quantum mechanics did not apply to those times?

In this discussion, we will concentrate on how the Copenhagen approach fits in with ours as a set of special cases and how the ``classical world'' can be replaced by  a quasiclassical realm. Such a realm is not {\it postulated} but rather is {\it explained} as an emergent feature of the universe characterized by $H$, $|\Psi\rangle$, and the enormously long sequences of accidents (outcomes of chance events) that constitute the coarse-grained decoherent histories. The material in the preceding sections can be regarded as a discussion of how  quasiclassical realms emerge.

We say that a `measurement situation' exists if some variables (including such quantum-mechanical variables as electron spin) come into high correlation with a quasiclassical realm.  In this connection we have often referred to fission tracks in mica. Fissionable impurities can undergo radioactive decay and produce fission tracks with randomly distributed  definite directions. The tracks are there irrespective of the presence of an ``observer''. It makes no difference if a physicist or other human or a chinchilla or a cockroach looks at the tracks. Decoherence of the alternative tracks induced by interaction with the other variables in the universe is what allows tracks to exist independent of ``observation'' by an ``observer''. All those other variables are effectively doing the ``observing''. . The same is true of the successive positions of the Moon in its orbit  not depending on the presence of ``observers'' and for density fluctuations in the early universe existing when there were no observers around to measure them.

The idea of ``collapse of the wave function'' corresponds to the notion of variables coming into high correlation with a quasiclassical realm, with its decoherent histories that give true probabilities. The relevant histories are defined only through the projections that occur in the expressions for these probabilities [cf \eqref{twoone}]. Without projections, there are no questions and no probabilities.   In many cases conditional probabilities are of interest.  The collapse of the probabilities  that occurs in their construction is no different from the collapse that occurs at a horse race when a particular horse wins and future probabilities for further races conditioned on that event become relevant.

The so-called ``second law of evolution'',  in which a state is `reduced' by the action of a projection, and the probabilities renormalized to give ones conditioned on that projection, is thus not some mysterious feature of the measurement process.  Rather it is a natural consequence of the quantum mechanics of decoherent histories, dealing with alternatives  much more general than mere measurement outcomes. 

There is thus no actual conflict between the Copenhagen formulation of quantum theory and the more general quantum mechanics of closed systems DHQM. Copenhagen quantum theory is an approximation to the more general theory that is appropriate for the special case of measurement situations. DHQM rather is a {\it generalization} of CQM.
That connection means that the experimental successes of CQM support DHQM as well. 

In our opinion decoherent histories quantum theory advances our understanding in the following ways among many others: 

\begin{itemize}
\item Decoherent histories quantum mechanics extends the domain of applicability of quantum theory to histories of features of the universe irrespective of whether they are receiving attention of observers and in particular to histories describing the evolution of the universe in cosmology.  
 
\item The place of classical physics in a quantum universe is correctly understood as a property of a particular class of sets of  decoherent coarse-grained alternative histories --- the quasiclassical realms \cite{Jer,qcrealm}.
In particular, the {\it limits} of a quasiclassical description can be explored. Dechoherence may fail if the graining is too fine. Predictability is limited by quantum noise and by the major branchings that arise from the amplification of quantum phenomena as in a measurement situation. Finally, we cannot expect a quasiclassical description of the universe in its earliest moments where the very geometry of spacetime may be undergoing large quantum fluctuations. 

\item Decoherent histories quantum mechanics provides new connections such as the relation   between the coarse graining characterizing quasiclassical realms and the coarse graining characterizing the  usual thermodynamic entropy of chemistry and physics. 

\item Decoherent histories quantum theory helps with understanding  the Copenhagen approximation. For example, measurement was characterized as an ``irreversible act of amplification'', ``the creation of a record'', or as ``a connection with macroscopic variables''. But these were inevitably imprecise ideas. How much did the entropy have to increase, how long did the record have to last, what exactly was meant by ``macroscopic''? Making these ideas precise was a central problem for a theory in  which measurement is fundamental. But it is less central in a theory where measurements are just special, approximate situations among many others. 
Then characterizations such as those above are not false, but true in an approximation that need not be exactly defined.

\item Irreversibility clearly plays an important role in science as illustrated here by the two famous applications to quantum-mechanical measurement situations and to thermodynamics. It is not an absolute concept but context-dependent like so much else in quantum mechanics and statistical mechanics. It is highly  dependent on coarse graining, as in the case of the document shredding. This was typically carried out in one dimension until the seizure by Iranian ``students'' of the U.S. Embassy in Tehran in 1979, when classified documents were put together and published. Very soon, in many parts of the world, there was a switch to two-dimensional shredding, which still appears to be secure today.
 It would now be labeled as irreversible just as the  one-dimensional one was previously. The shredding and mixing of shreds clearly increased the entropy of the documents, in both cases by an amount dependent on the coarse grainings involved. Irreversibility is not absolute but dependent on the effort or cost involved in reversal.

\end{itemize}. 
\section{Acknowledgements}
The author's understanding  of quantum theory has been deeply influenced by discussions with the late Murray Gell-Mann, Thomas Hertog, and Mark Srednicki  over a long period of time. He thanks the Santa Fe Institute for supporting many productive visits there. This work was supported in part by the National Science Foundation under grant PHY15-04541 and PHY18-18018105 .

\end{document}